\documentclass[pra,superscriptaddress,10pt,a4paper,showpacs]{revtex4}
\usepackage{graphicx,amsmath,amsfonts,algorithm}
\usepackage{graphics}
\usepackage{amssymb}
\usepackage{enumitem}
\usepackage{color}
\usepackage{bbm}

\def\ket#1{| #1 \rangle}

\def\ket#1{\left| #1 \right\rangle}
\def\bra#1{\left\langle #1 \right|}
\newtheorem{prop}{Proposition}\def\PRO{\begin{prop}}\def\ORP{\end{prop}}
\newtheorem{coro}{Corollary}\def\COR{\begin{coro}}\def\ROC{\end{coro}}
\newtheorem{theo}{Theorem}\def\TH{\begin{theo}}\def\HT{\end{theo}}
\def\TH{\begin{theo}}\def\HT{\end{theo}}
\newtheorem{defi}[prop]{Definition}\def\DE{\begin{defi}}\def\ED{\end{defi}}
\newtheorem{lemme}[prop]{Lemma}\def\LE{\begin{lemme}}\def\EL{\end{lemme}}

\def\EQ#1{\begin{eqnarray}#1\end{eqnarray}}

\def\op#1{\hat{#1}}

\def\op#1#2{|#1\rangle\langle#2|}

\def\dm#1{\op{#1}{#1}}

\newcommand{\djj}{d\kern-0.4em\char"16\kern-0.1em}
\def\sect#1{{\it #1}.---}

\begin{document}
 \title{Quantum Digital Signatures without quantum memory}
\author{Vedran Dunjko}
\email{vdunjko@inf.ed.ac.uk}
\affiliation{School of Informatics, University of Edinburgh, Edinburgh EH8 9AB, UK}
\affiliation{Division of Molecular Biology, Ru\djj er Bo\v{s}kovi\'{c} Institute, Bijeni\v{c}ka cesta 54, P.P. 180, 10002 Zagreb, Croatia}
\affiliation{SUPA, School of Engineering and Physical Sciences, Heriot-Watt University, Edinburgh EH14 1AS, UK}

\author{Petros Wallden}
\email{petros.wallden@hw.ac.uk}
\affiliation{SUPA, School of Engineering and Physical Sciences, Heriot-Watt University, Edinburgh EH14 1AS, UK}
\affiliation{Physics Department, University of Athens, Panepistimiopolis 157-71, Ilisia Athens, Greece}
\author{Erika Andersson}
\email{e.andersson@hw.ac.uk}
\affiliation{SUPA, School of Engineering and Physical Sciences, Heriot-Watt University, Edinburgh EH14 1AS, UK}

\begin{abstract}
Quantum Digital Signatures (QDS) allow for the exchange of messages from one sender to multiple recipients, with the guarantee that messages cannot be forged or tampered with. Additionally, messages cannot be repudiated -- if one recipient accepts a message, she is guaranteed that others will accept the same message as well. While messaging with these types of security guarantees are routinely performed in the modern digital world, current technologies only offer security under computational assumptions. QDS, on the other hand, offer security guaranteed by quantum mechanics. All thus far proposed variants of QDS require long-term, high quality storage of quantum information, making them unfeasible in the foreseeable future. Here, we present the first QDS scheme where no quantum memory is required, and all quantum information processing can be performed using just linear optics. This makes QDS feasible with current technology.
\end{abstract}

\maketitle
\sect{Introduction}
Quantum Digital Signatures (QDS), originally proposed in \cite{QDS}, is an unconditionally secure cryptographic primitive for the exchange of classical messages between one sender and multiple receivers with two types of security guarantees: \emph{security against forging} and \emph{security against repudiation}.
Security against forging guarantees that none of the recipients (or other parties) can forge a message, or alter a message which is sent from the sender to one of the recipients. Security against repudiation (sometimes called \emph{transferability} \cite{QDS}) implies that a malevolent sender cannot cause two (or more) recipients to disagree on the validity of a message.
Currently, these two types of security guarantees for message sending are achieved using classical public-key based digital signatures (DS) protocols, and are constantly
required in modern communication.
The key advantage of QDS over traditionally used schemes is in the flavor of security -- DS offer security under computational assumptions~\footnote{These assumptions are often  unproven statements of computational hardness of certain mathematical problems.} whereas the security in QDS holds against computationally unbounded adversaries.

In a generic QDS protocol the sender sends (a bounded number of) copies of pairs quantum states to the multiple receivers. These states, which we will refer to as \emph{quantum signatures}, are then kept in quantum memory by the recipients, until the sender decides to send a particular message. Authenticity is effectively guaranteed by the fact that a limited amount of information in the emitted quantum signatures is accessible to potential forgers.
Non-repudiability is enforced by having the recipients perform some type of non-destructive quantum state comparison on the initially sent quantum signatures (for instance, a SWAP test \cite{QDS}).
Although general non-destructive state comparison is currently beyond
experimental limits, in \cite{ErikaOrig} a QDS scheme based on coherent states was proposed, where the comparison can be performed using simple linear optics.
This scheme was recently implemented experimentally~\cite{Expr}. However, the remaining and more challenging requirement for viable QDS
is the quantum memory, which may have to store millions of qubits (or qumodes), coherently, perhaps for months or years. This is a substantially impractical requirement given that
state-of-the-art memories cannot achieve coherence times beyond the scale of minutes \cite{qmem}.

In this Paper we circumvent the requirement of quantum memory. We propose a
QDS scheme with the same security guarantees as those in \cite{QDS, ErikaOrig, Expr}, but without the need of quantum memory. The scheme can be implemented using just linear optics and photodetectors that distinguish only between zero and non-zero photons.

\sect{QDS without quantum memory}
QDS protocols can be separated into two stages: the 
\emph{distribution stage}, where quantum signatures are sent to all future recipients,
and the \emph{messaging stage}, where classical messages are sent and verified. In its simplest form, the distribution stage enables a sender (\textit{Alice}) to send a message
to either or both of two recipients (\textit{Bob} and \textit{Charlie}) at some point in the future (during the
messaging stage). The distribution stage is independent of the future message sent in the messaging stage. The main difference between our protocol and previous protocols is in the first stage. While existing protocols end with the recipients storing quantum states in quantum memory for future use, in our protocol, those states are immediately measured. Thus, only classical information has to be stored. We will show that the classical measurement outcomes at the end of the distribution stage can still be used for
secure distribution of messages, thus eliminating the need for quantum memory.

\begin{figure}
\includegraphics[trim=170 170 100 100, clip,width=8cm,height= 8 cm]{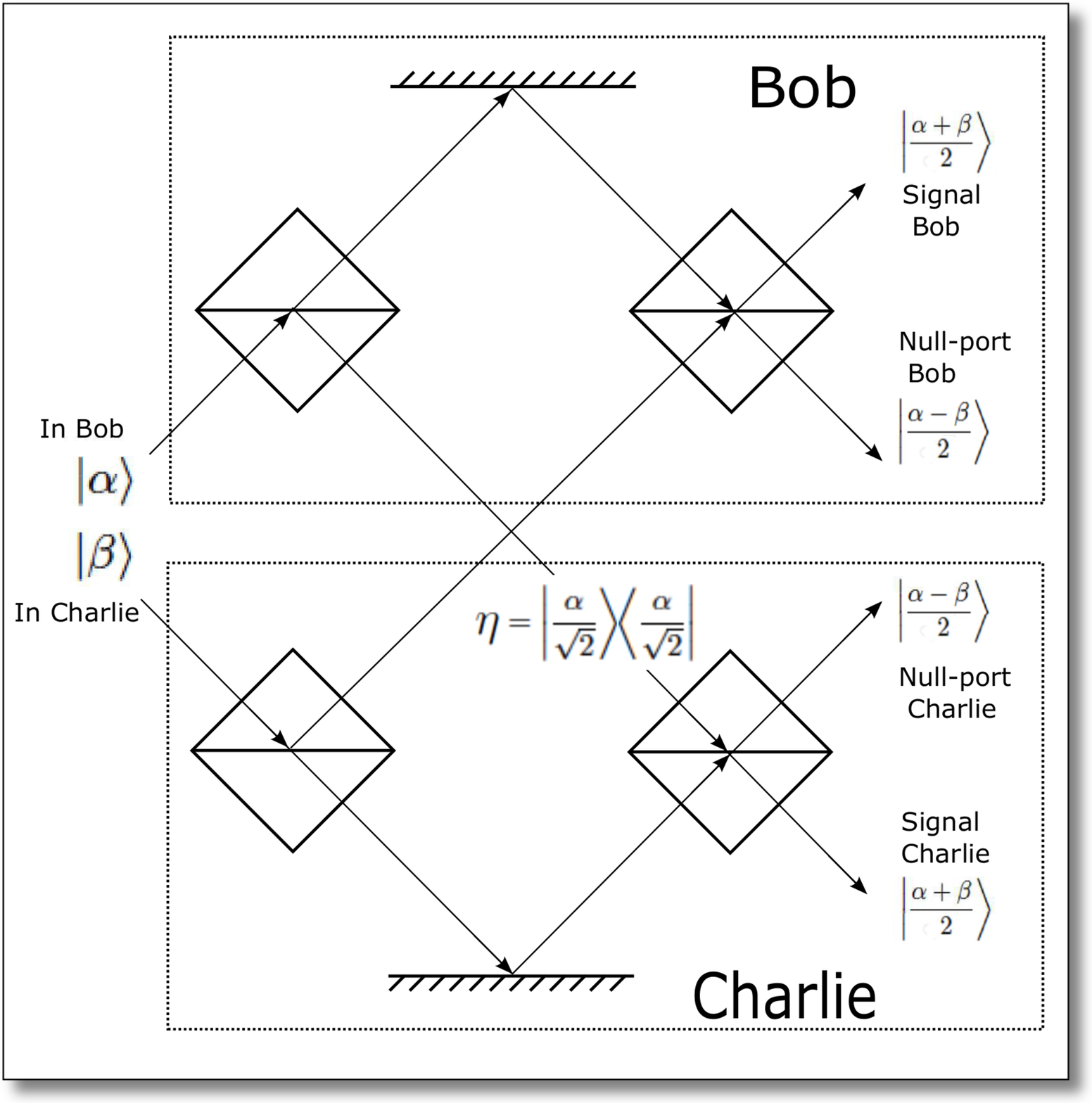}
\caption{\label{Fig:multi}The multiport: The out-signal arms contain a state symmetric under a swap.
The null-ports contain the vacuum if and only if the in-signal arms contained a symmetric state.  In active cheating Bob can choose what state $\eta$ he sends back to Charlie, so as to
optimize his cheating strategies. However, Charlie's null-port counts measure the fidelity between honest and dishonest states $\eta$. In the figure, $\eta$ is set to the honest response state.}
\end{figure}

Our QDS protocol is based on the QDS scheme proposed in \cite{ErikaOrig}.  For simplicity, we will  consider the case with two receivers, but will explain how this can be generalized later. The quantum signatures comprise trains of coherent states randomly chosen by Alice as $|\alpha\rangle$ or $|-\alpha\rangle$.
In the scheme we present here, just as in~\cite{ErikaOrig, Expr}, non-repudiability is ensured by the use of a multiport, shown in Figure \ref{Fig:multi}.
The multiport  is a passive linear optical device comprising four 50:50 beam splitters .
The top two belong to Bob and the bottom two to Charlie.
The input states to both Bob's and Charlie's first beam splitters are a vacuum state and the inbound state from Alice. The outputs of the first two beam splitters comprise the inputs of the second two beams splitters, as shown in Fig. \ref{Fig:multi}.
 We will refer to the output ports of the second two beam splitters as Bob's and Charlie's signal port and null-port, respectively.

Intuitively, the multiport non-destructively  \footnote{The action of the multiport is non-destructive provided that the inbound states were actually the same, which happens, for instance, when Alice was honest.} compares the coherent states entering at Bob's and Charlie's in-ports.
Non-repudiability is, however, guaranteed in a somewhat different way. There are two properties which make the multiport central for the security of the QDS scheme. Firstly, it symmetrizes inbound states.
More precisely, for any input state, the joint state of Bob's and Charlie's quantum signatures,
exiting from their signal ports, is invariant under swap of every matching quantum signature element pair (see Appendix \ref{Supp} for details).
This will guarantee
security against repudiation, even without considering the counts on the null-ports, that is, without actively checking whether Alice sends in identical states.
Secondly, the null-port counts can serve as a safeguard against forgers who act maliciously in the distribution stage as we will shortly describe. Similar to all other QDS proposals, we assume that player-to-player classical communication is performed over an authenticated (but not confidential) classical channel, and that the quantum channel from Alice to the (multiport in-ports of) the recipients is authenticated as well.
The classical authenticated channels can easily be realized using a classical message authentication scheme with a small overhead of pre-shared keys. The requirement for the quantum authenticated channel we address in the concluding discussion.

At the end of the distribution stage,  the received quantum signatures are measured using  unambiguous state discrimination (USD)
\cite{Ivanovic,Dieks,Peres}, which for the case of the two coherent states of opposite phase can be optimally realized using linear optics alone \cite{Banaszek}.
An unambiguous quantum measurement gives a result that is (in the ideal case) guaranteed to be correct, at the expense of sometimes failing to give a result at all.
In the subsequent messaging stage, Alice accompanies a message with the information about what signs she chose for the corresponding sequence of coherent states, and the recipients verify that a low enough number of signs -- the declared phases of the coherent states -- fail to match those measured during the distribution stage.

A more specific protocol outline is given next.
The basic protocol describes how single-bit messages can be securely distributed. For longer messages, the distribution stage of the protocol is simply iterated, and all bits of the message are later checked in the also iterated messaging stage.
We will introduce a number of \textit{internal parameters} which have to be appropriately chosen.
We also introduce an \textit{external parameter} $L$,
the choice of which directly influences the security levels of our scheme. Finally, we introduce
$p_{USD} = 1-e^{-2 \alpha^2},$
which is the optimal success probability of an unambiguous discrimination of the states $\{\ket{\alpha}, \ket{-\alpha} \}$ \cite{Ivanovic,Dieks,Peres,Banaszek}.
  \begin{itemize}
  \setlength{\itemindent}{-.5cm}
 \item[] \textit{Distribution stage}
 \begin{enumerate}[leftmargin=-.3cm]
\item For each possible future message $k=0,1$, Alice generates two copies of a sequence of coherent states (called \emph{quantum signatures})
$QuantSig_k = \bigotimes_{l=1}^L \rho_l^k$
where
$\rho_l^k = \dm{b_l^k \alpha}$, $\alpha$ is a real positive amplitude, $b_l^k \in \{-1,1 \}$  are randomly chosen signs, and $L$ is a suitably chosen integer.
The state $QuantSig_k$ and the sequence of signs $PrivKey_k = (b_1^k, \ldots b_L^k)$ are called the \textit{quantum signature} and the \textit{private key}, respectively, for message $k$.
The individual state $\rho_{l}^k$ we refer to as the $l^{th}$ \emph{quantum signature element} state for
message $k$.

\item Alice sends one copy of $QuantSig_k$ to Bob and one to Charlie,
for each possible message $k=0$ and $k=1$.

\item Bob and Charlie send their sequences $QuantSig_k$ for both $k=0$ and $k=1$, one signature element at a time, through the QDS multiport, shown in Fig. \ref{Fig:multi}. For each signature element they note whether photons are registered at their multiport null-port.

\item \label{prot:USD}
Bob and Charlie each measure all the multiport signal states using the USD measurement for $\left\{ \ket{\alpha}, \ket{-\alpha}\right\}$. They store the unambiguous outcomes (along with the index of the state for which it occurred) for  $k=0,1$.
Thus they obtain and store triplets of the form $\{  (k, l' , b^k_{l'})      \}$ where $1\le l'\le L$.
 \end{enumerate}

\item[] \textit{Messaging stage}
\begin{enumerate}[leftmargin=-.3cm]
\item For the bit message $m$, Alice sends $(m,PrivKey_m)$ to the desired recipient (say Bob).
\item To authenticate, Bob checks whether $(m,PrivKey_m)$ matches with his measured stored sequence, for the positions where he obtained the unambiguous outcome. He authenticates if the number of mismatches is below $s_a p_{USD} L$, where $s_a$ is an \textbf{authentication threshold}.
\item Before accepting the message, Bob confirms he has no reason to abort the protocol, by checking the following two things. (a) If the number of signature elements for which non-zero null-port counts are registered breaches a threshold $r L$ for $0\leq r <1$ he aborts. (b) If the number of unambiguous outcomes is not inside the expected interval $[ (p_{USD} - \delta) L,(p_{USD} + \delta) L]$, where $0<\delta<1$ is called the \textbf{unambiguous count tolerance}, he aborts.
\item To forward the message to Charlie, Bob forwards to
Charlie the pair  $(m, PrivKey_m)$ he received from Alice. Charlie then performs an
analogous procedure to Bob, where he verifies the message if the number of mismatches  is below $s_v p_{USD} L$ where $s_v$ is called
the \textbf{verification threshold}, with $0 \leq s_a < s_v <1$.
\item For Charlie to accept the message, he also confirms that he has no reason to abort the protocol, by repeating the steps Bob did above.
\end{enumerate}
\end{itemize}
In the description above, the roles Bob and Charlie play are arbitrary. One player performs the authentication procedure, whereas the other the verification of the forwarded message.

\sect{Protocol performance}
We will be interested in three performance properties of this protocol: \emph{correctness},  \emph{security against repudiation} and \emph{security against forging}.
Another important property, \emph{robustness}, which guarantees that the protocol will work even if physical imperfections are considered, we address briefly later in this Paper, and leave a more thorough analysis for future work.
Below we will show that the probabilities of incorrect behavior, forging and repudiation events decay exponentially in $L$, for an appropriate choice of the internal parameters. 
We will also show that there exists a consistent assignment of these parameters which yields an exponential decay of the probabilities of all the undesired events. Note that in the three-player setting it only makes sense to consider scenarios where at most one player is malevolent, since two or more malevolent players can always trivially cheat on the third.

Correctness:
Correctness for our protocol implies that if all parties behave honestly (and no imperfections are present), then the protocol is aborted only with a negligible probability, and the message is accepted by the recipients.
In our protocol an honest case abort can only occur if either the number of  unambiguous outcomes either of the parties obtains is outside the tolerance window $[p_{USD} - \delta, p_{USD} + \delta ]$. This occurs with probability
\EQ{
P(honest\ abort)  \leq (1-2\exp \left(-2\delta^2 L \right))^2. \label{correctness}
}
To see this, note that the expected value for the number of unambiguous outcomes, for each player and in the honest case, is exactly $p_{USD}L$. The expression $2\exp \left(-2\delta^2 L \right)$, by Hoeffding's inequalities \cite{hoeff},  bounds the probability that the deviation from this mean is larger than $\delta L$, and the overall expression takes into account that neither player should abort.

Security against repudiation:
Repudiation occurs when the protocol is not aborted, the message is authenticated by one of the parties, but the same message gets rejected when forwarded to an another party.
Thus, in this setting, Alice is the malevolent player, and her strategies effectively comprise the possible choices of quantum states she sends in the distribution stage.

Security against repudiation relies on the symmetrization property of the multiport -- the joint states of Bob's and Charlie's signal ports are invariant under swaps of matching quantum signature elements, and the fact that the acceptance thresholds $s_a$ and $s_v$ differ, so that  $s_v>s_a$.
The strong symmetrization property of the multiport implies, for instance, that the matching reduced density matrices for Bob and Charlie are equal for each key element.
That alone will guarantee the security against individual attacks. However,
also when post-selecting on some sequence of outcomes for Bob and Charlie, for part of the quantum signature elements, the unmeasured remainder of the signatures is still symmetric under swap of matching elements. The latter property guarantees security against coherent attacks (see Appendix \ref{Supp} for details).

Intuitively, any strategy of Alice, besides the honest one, will yield a certain average fraction of mismatches (a fraction Alice can control) between the declared private key, and the signs
measured using the USD. Since Bob's and Charlie's reduced states are equal, and the declarations they compare their measurement outcomes with are equal, the fraction of mismatches will, on average, be equal for both players.
So, it is not possible for Alice to deterministically both make Bob observe less than $s_a p_{USD}L$ mismatches and Charlie to observe more than $s_v p_{USD} L$ mismatches. Alice's only freedom is to choose the average number of mismatches, which can occur for either Bob and Charlie with equal probabilities. It can be shown that her optimal choice
is $(s_v-s_a)p_{USD}/2$. In this case, the bound on the probability of Alice successfully repudiating is given by

\EQ{
P(repudiation) \leq \exp \left(- \dfrac{1}{2} p_{USD}^2(s_v - s_a)^2 L\right). \label{repudiation}
}
Eq. (\ref{repudiation}) again stems from Hoeffding's inequalities, which give the probability that the number of matches deviates from the expectancy by more than a fraction, as a function of $L$.
The full proof of this claim is more involved and can be found in Appendix \ref{Supp}.
It also shows that using classical correlations or entanglement does not help Alice.

Security against forging:
By forging we denote the scenario where a dishonest recipient convinces an honest recipient that Alice has sent a message  when Alice has sent no message at all (message forging), or a message $m'$ differing from the message $m$ Alice has in fact sent (message tampering).
For our setting, the probability of successful message tampering equals the probability of message forging, as the private keys for two differing messages are independently distributed. Without loss of generality, we will assume that the forging party is Bob.

We identify different types of forging attacks. Firstly, we distinguish between \emph{passive} and \emph{active} attacks. In passive attacks, Bob behaves honestly in the distribution stage until step \ref{prot:USD} of the protocol. Here, he stores all the quantum systems outbound from the multiport in quantum memory, and performs measurements which will optimize his cheating probability. In active attacks, Bob acts maliciously throughout the distribution stage. Specifically, Bob can tamper with his part of the multiport.

Secondly, we distinguish between \emph{individual}, \emph{collective} and \emph{coherent} attacks. This classification is reminiscent of
the traditionally studied attacks in quantum key distribution (QKD) \cite{Attacks}. In individual attacks, Bob's actions (whether it is measuring the quantum signature or tampering with his section of the multiport) are independent for each signature element.
In collective attacks, he may use only strategies which are classically correlated for different signature elements.
For coherent attacks we remove this last constraint, allowing quantum correlations.
Such attacks constitute the most general type of forging activity.
Here, we will address the security of our protocol both for passive and active attacks, for individual and collective attacks, and leave the analysis of coherent attacks for future work.
Intuitively, since different signature elements are independent, Bob should have nothing to gain from quantum correlations.

In passive attacks,
Bob's goal is to make Charlie accept
$PrivKey_{m'}$ for the single bit message $m'$ of his choice. Note that knowing
$PrivKey_{m' \oplus 1}$ for message $m'\oplus 1$
does not help, since the signs of the two messages are independently distributed.
Similarly, since all the signs within one private key are independently distributed, the optimal collective forging
strategy can be shown to be (see Appendix \ref{Supp})  performing minimum-error measurements \cite{Helstrom} on each of the signature elements
in $QuantSig_{m'}$.
The results of the minimum-error measurements are reported to Charlie who then checks them against his unambiguous outcomes. Let $p_{min}$ be the minimum-error probability, \emph{i.e.} the probability that Bob incorrectly identifies a quantum signature element, and let $p_{USD}$ be the probability of obtaining the unambiguous outcome in the USD measurement. For the case of two states it holds that~\cite{Ivanovic,Dieks,Peres}
\EQ{
p_{min}&=&1-\frac{1}{2} \left(\sqrt{1-e^{-4 \alpha^2}}+1\right).
}
For Bob to successfully forge with Charlie (that is, have a faked message verified, which is easier than to forge a transferrable message as $s_a < s_v$), Bob must correctly guess a fraction of the unambiguous measurement outcomes  Charlie obtained. Since forging by definition can occur only if Charlie did not abort, Charlie has received at least $p_{USD}'L$  unambiguous outcomes, for $p_{USD}' = p_{USD}-\delta$. This lower bound is also the best scenario for forger Bob, and for this case one can derive the bound
\EQ{
P(forge) \leq \exp \left(-2  \left(p_{min} - s_v \dfrac{p_{USD}}{p_{USD}' } \right)^2 p_{USD}' L  \right). \label{forging}
}
By setting $\delta = 0$, Eq. (\ref{forging}) would bound the probability of Bob making an error in his estimate of the encoded phases fewer than $s_v p_{USD}L$ times, out of $p_{USD}L$ guesses.

In active individual attacks, Bob can prepare response states, denoted $\eta$, shown in Fig. \ref{Fig:multi}, which modifies Charlie's quantum signature states.
To counteract such attacks, Charlie must keep track of the multiport null-port counts
in the distribution stage, as they measure the effective fidelity between the passive-strategy response state and an active one (see Appendix \ref{Supp}). Requiring that these output ports
yield no detection events (\emph{i.e.} setting $r=0$) implies, in the asymptotic limit of infinite signature states $L \rightarrow \infty$,
 that Bob must have been honest throughout the distribution stage.
This reduces
active individual attacks to passive attacks.
For collective attacks, it is easy to see that the space of collective strategies forms a convex structure, and the optimal cheating probability is achieved at an extremal point -- which corresponds to an individual attack. Thus collective strategies are no better than individual. However, this reasoning does not simply extend to coherent attacks, which we leave for future work. We provide more details, and quantitative security statements
in Appendix \ref{Supp}.

Robustness and parameter constraints:
Equations (\ref{correctness}),(\ref{repudiation}) and (\ref{forging}) constrain the internal parameters to ensure  exponential decay in the probability of an unwanted event, as a function of $L$. Concretely, as long as $\delta>0$ and $s_v <p_{min}({p_{USD} - \delta}){p_{USD}} p_{min}$ (say $\delta = p_{USD}/10$ and $s_v = p_{min}/4$) we get the desired exponential decay.
The parameters $r$ and $s_a$ can in the ideal case be set to $0$. We have however left these parameters in the expressions and the protocol definition, as they can be used to counteract imperfections which will occur in any realization. For instance, if the USD procedure is not perfect, there
may be a mismatch even if everybody is honest, and this can be ameliorated by having a non-zero $s_a$. Similarly, multiport null-port counts can occur even if everybody is honest, for instance through detector dark counts. This can be counteracted by setting $r>0.$ However, the detailed analyses of the imperfect settings are beyond the scope of this Paper. See Appendix ~\ref{Supp} for an analysis taking into account slight modifications due to active forging.

We have assumed that all the players share a common trusted reference frame, which is necessary for the
phase of the coherent states to be defined. This could be realized by having a fourth party send sequences of strong reference pulses of a known amplitude, defining the phase $0$. Alternatively, Alice could send time-multiplexed pairs of signal-idler coherent states, where the idler serves as the reference beam, as was implemented in \cite{Expr}. In this case, the security analysis would be slightly different (as the reference beam could be tampered with) but with minor modifications our results would
still hold. The full analysis of this we also leave for future work.

\sect{Discussion}
Our protocol is easily generalized in many ways. First, using a generalized version of the multiport \cite{ErikaOrig} it is easily extended to any number of recipients.
Moreover, the quantum signatures could be chosen from more than two possible quantum states,
for instance the states $\ket{e^{i \theta} \alpha}$ with $\theta = 2 k \pi/N, k=0, \ldots N-1.$ For
$N>2$, unambiguous discrimination cannot be performed optimally using linear optics (except in the asymptotic limit). Non-optimal schemes, or asymptotically optimal schemes, such as proposed in \cite{vanEnk, UsAmplif, ExpUSD}, or state elimination schemes may however be good enough. It remains an open and highly non-trivial question, even in the ideal case, which choice of $\alpha$ and $N$ maximizes the security level for given
key lengths.

More generally, we can envisage qubit-based QDS schemes, or schemes based on linearly dependent sets of states,
where USD is not possible~\cite{CheflesImpos}.
For example, it would be interesting to investigate
schemes using the BB84 states, using minimum-error measurements or quantum state elimination.
A remaining issue, aside from the security proof against forging in the most general types of attacks, is the requirement that Alice has an authenticated quantum channel to each of the recipients.
While general quantum message authentication protocols \cite{AQM} are resource-expensive (unlike the classical counterparts),  in our case, verification of only two possible input states is needed.
Thus, potentially, techniques similar to those used in standard QKD could be employed to verify the channel (by sacrificing a fraction of the states).

Finally, we note that this protocol, since all the quantum signatures are immediately measured, highlights that certain types of classical multi-party correlations are sufficient for
secure digital signature schemes.
It is an interesting question which types of correlations are sufficient and
whether they can be achieved by means other than a QDS-type distribution stage (for instance by many  point-to-point QKD systems).

Support by EPSRC grants No. EP/G009821/1, EP/K022717/1, the EPSRC Doctoral Prize scheme, and
partial support from COST Action MP1006 is gratefully acknowledged.
\newpage

\section{Appendix: Supplementary Material}\label{Supp}

\subsection{Outline and Definitions}

In this supplementary material we will in detail examine the performance of the protocol, presented in the main paper. We will prove security against repudiation for any type of attack. As for security
against forging, we will prove the security against all collective active attacks, as defined in the main paper. Recall, in general there are three types of attacks (for both repudiation and forging) that a malevolent party may use, of increasing generality. The first are the individual attacks, where the attack happens independently (but not necessarily identically) to each element of the quantum signature. Second are the collective attacks, where the malevolent party may use (classical) correlations between the different elements of the quantum signature, but no quantum correlations. Finally, the general coherent attacks allow the malevolent party to use any possible attack strategy, including entangled correlations between different elements of the quantum signature.

The outline of this document is as follows. We will first give relevant definitions. We will then prove certain properties of the multiport, for use in the remainder of the document. Following this we will examine the security against repudiation, against forging and finally review the correctness of the protocol and give a suitable choice for the parameters which appear in the protocol.

To remind the reader, the protocol is divided in two stages: the distribution and the messaging stage. During the distribution stage, Alice sends quantum signatures  $\otimes_{i=1}^L\ket{b_i\alpha}$, composed of $L$ coherent states, to Bob and Charlie. The $b_i$ are chosen uniformly at random in $\{-1,1 \}$, independently for each of the two possible messages $m=0,1$. Thus each coherent state in the quantum signature is either $|\alpha\rangle$ or $|-\alpha\rangle$.
By convention, we assume $\alpha$ is real and non-negative.
Bob and Charlie pass the quantum states they receive from Alice
through the multiport, and they measure, using the USD measurement, the resulting state from the signal-port, while they also keep track of the null-port counts to protect from active forging, as we will see later. We will call the sequence of their outcomes {\it measured signatures}. The probability for obtaining an unambiguous outcome depends on the amplitude of the coherent states and is given by

\EQ{p_{USD}=1-e^{-2\alpha^2}.}

During the messaging stage, Alice sends, for the desired message $m \in \{0,1 \}$, the so-called private key $PrivKey_m$ which is the pair of the message $m$ and the classical string $\{b_i\}^L_{i=1}$ to Bob. Here $b_i \alpha$ defines the coherent state at the $i^{th}$ position in the quantum signature matching the message $m$. We will call the classical string defining the phases of the states \emph{Alice's declaration} in the remainder of this paper.
As is clear from the protocol definition, the actual phases, measured or declared, do not directly determine whether Charlie or Bob accept or reject the message. It is the match or mismatch between the two which solely determines the decisions of Bob and Charlie.
Thus, it will be convenient to define as $+1$ the event when a measurement by Bob or Charlie
matches Alice's declaration, and as $-1$ when there is a mismatch. If
the USD measurement failed to produce an outcome, we will denote this event $0$.
This is equivalent to noting that, since only the match or mismatch matter, we can without loss of generality assume that Alice declares a string of $+1$ phases, so that the events $-1,1,0$ defined above coincide exactly with the actual phase measurement outcomes, with 0 denoting the ambiguous outcome.

As stated, Bob authenticates  the message (an event we will call ``BA'') depending on how well
his measured signatures matches Alice declarations. More specifically, he authenticates if he finds less than $s_a p_{USD}L$ mismatches ($-1$'s). Here, $0 \leq s_a <1$ is an internal parameter of the the protocol.

Subsequently, Bob forwards Alice's declaration to Charlie, or in the case of forging he forwards his own declarations which supposedly came from Alice. Charlie verifies (event ``CV'') that the message came from Alice, if he finds less than $s_v p_{USD}L$ mismatches, where $s_v$ is an internal parameter which has to satisfy $s_a < s_v <1$. If Charlie finds more mismatches then he rejects (event ``CR'').
Note here that we choose Bob to be the first party to receive the declaration in order to authenticate, while Charlie is the party receiving the forwarded message and verifying it. This choice is arbitrary.

The protocol is aborted in two different cases.
 Firstly, abort occurs if either Bob or Charlie obtain too many or too few unambiguous outcomes in the distribution stage.
More precisely, abort occurs  if the total number of observed unambiguous outcomes (for either Bob or Charlie) is outide the interval $[(p_{USD}-\delta)L, (p_{USD}+\delta)L]$ \footnote{Note that $p_{USD}L$ is exactly the expected number of unambiguous outcomes.}, where $\delta$ is called the unambiguous count tolerance and will be suitably chosen later. Secondly, abort occurs if either Bob or Charlie observe more than $r L$ photon detection events at the null-ports of the multiport, again in the distribution stage.
This could mean that the quantum signatures received from Alice were not identical, or, as we will clarify later, that Bob attempted to forge actively.  Note however, that the null-port abort case will not be required to prove security against repudiation and only appears in the analysis of active forging attempts.
  We will call the event in which Bob does not abort ``BNA'', while the event Charlie does not abort ``CNA''.

For the protocol to be correct we will require that Bob authenticates, Charlie verifies and neither aborts if all parties are honest, except with negligible probability. In particular we will require that the probability  \EQ{P(correct)=P(BA,CV,BNA,CNA),} tends to unity exponentially quickly in the quantum signature length $L$.

Concerning security statements, we will deem the protocol secure against repudiation if the probability of  Bob authenticating, Charlie rejecting, and neither of them aborting
occurs only with exponentially small probability in $L$ for any attack strategy by Alice.
We will thus need to show that \EQ{P(rep)=P(BA,CR,BNA,CNA)} decays exponentially quickly in $L$.
 In the proof of security against repudiation, which we present here, we will actually bound the probability $P(BA,CR)$ by an exponentially decaying value, but this will then bound $P(rep)$ as well, given that $P(BA,CR,BNA,CNA) \leq P(BA,CR)$.

 Forging occurs if Bob sends his declaration, without having access to the declaration of Alice, and subsequently Charlie verifies, and neither party aborts. Thus we will require that the probability of such an event, that is, \EQ{P(forge)=P(CV,BNA,CNA),} again decays exponentially quickly in $L$.
In our proofs we will actually bound  $P(CV,BNA|CNA),$ which is a bound for $P(forge)$ as well, since $P(CV,BNA,CNA)\leq P(CV,BNA|CNA)$.

Note that our definitions concern one isolated run of the protocol, which is a flavor of security sometimes called \textit{stand-alone} security.  Stronger security claims,  such as composable security claims, we leave for future work.

We will also use the following forms of Hoeffding's inequalities \cite{hoeff}.

\LE Let  $X_1,\cdots, X_L$ be independent random variables, each attaining values $0$ or $1$. Let $\bar{X}=1/L\sum X_i$ be the empirical mean of the variables, and let $E(\bar{X})$ be the expected value of $\bar{X}$. Then for all $t\geq 0 $ we have

\EQ{P(\bar{X}-E(\bar{X})\geq t)\leq \exp (-2t^2 L)\label{Hoeffding 1}}

\EQ{P(E(\bar{X})-\bar{X}\geq t)\leq \exp (-2t^2 L)\label{Hoeffding 2}}

\EQ{P(|\bar{X}-E(\bar{X})|\geq t)\leq 2 \exp (-2t^2 L)\label{Hoeffding 3}.}
\EL
A final point to stress is that here, as in the main paper, we will deal with the ideal case, where there are no losses and imperfections in the apparatus. Dealing with the experimentally implementable set-up is ongoing work. We have however included some parameters that are absolutely necessary only for the non-ideal case.

\subsection{The multiport}

\begin{figure}
 \includegraphics[width=10cm,height= 10 cm]{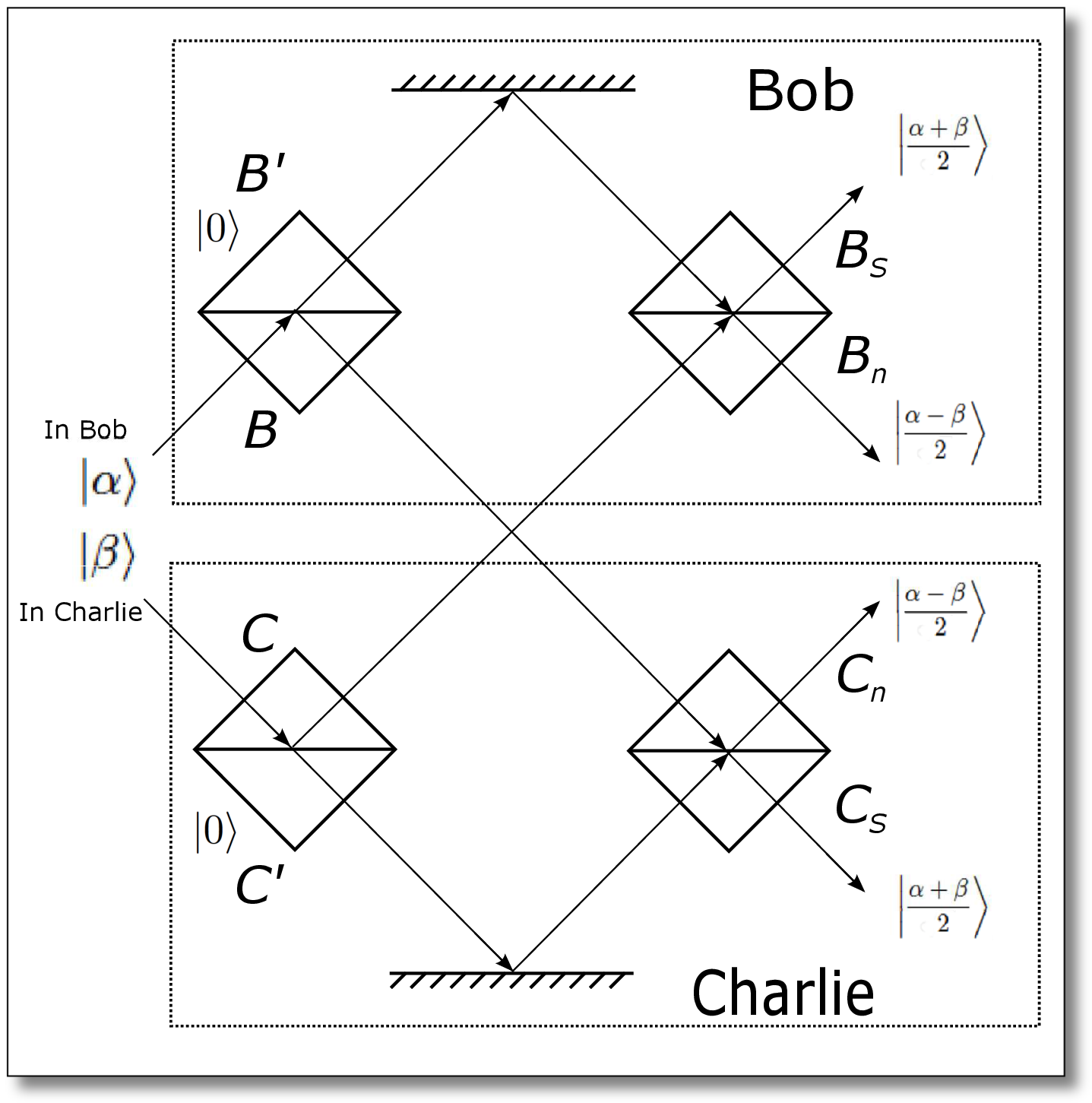}
\caption{\label{Fig:mult1} The multiport with the input-output relations given in terms of the coherent state basis.}
 \end{figure}

The multiport is a passive linear optical device with 4 input modes and 4 output modes, comprising four
50:50 beamsplitters. Two of the input modes always contain the vacuum state.
The  beam splitters we use act on the field operators according to
\EQ{
\binom{\hat{a}_{out}^{\dagger}}{\hat{b}_{out}^{\dagger}} = \dfrac{1}{\sqrt{2}} \left(
\begin{tabular}{cc}
1 & 1\\
1 & -1\\
\end{tabular}
 \right)
\binom{\hat{a}_{in}^{\dagger}}{\hat{b}_{in}^{\dagger}}.
}
 The  multiport and the input-output relations for coherent state inputs are illustrated
 in Figure \ref{Fig:mult1}.
The top two beam splitters are held by Bob and the lower two by Charlie.
Since two of the four input modes of the multiport are always set to the vacuum state by Bob and Charlie, we will refer to the remaining two modes as the input modes of the multiport.
For an input state where two of the input modes are in the vacuum state while the other two are coherent states $\alpha$ and $\beta$, the multiport acts according to
\EQ{\ket{\alpha}_B\otimes\ket{\beta}_C\otimes\ket{0}_{B'}\otimes\ket{0}_{C'}\rightarrow \ket{\frac{\alpha+\beta}2}_{B_s}\otimes\ket{\frac{\alpha+\beta}2}_{C_s}\otimes\ket{\frac{\alpha-\beta}2}_{B_n}\otimes\ket{\frac{\alpha-\beta}2}_{C_n}.}
Alice controls the (non-vacuum) input states of Bob and Charlie. We will assume that she can send the most general mixed entangled state of the form
\EQ{\rho_{BC}=\sum_i p_i\ket{\psi_i}_{BC}\bra{\psi_i}_{BC}.}
The state $\rho_{BC}$ refers to a single element pair (Bob and Charlie's corresponding pair) of the signature. Rewriting this state in terms of its purification, with the help of a (fictitious) environment, we can write it as a pure state in an extended Hilbert space,
 \EQ{\ket{\psi_{in}}_{BCE}=\sum (p_i)^{1/2} \ket{\psi_i}_{BC}\otimes\ket{i}_E}
 which can be rewritten again using the (scaled) P-representation as
 \EQ{\ket{\psi_{in}}_{BCE}=\int \sum_i P_i(\alpha,\beta)\ket{\alpha}_B\otimes\ket{\beta}_C\otimes\ket{i}_Ed\alpha d\beta.}
 The effect of the multiport on the total state (where we have included the two extra vacuum input states) is
 \EQ{\ket{\psi_{out}}_{B_s,C_s,B_n,C_n,E}=\int \sum_i P_i(\alpha,\beta)\ket{\frac{\alpha+\beta}2}_{B_s}\otimes\ket{\frac{\alpha+\beta}2}_{C_s}\otimes\ket{\frac{\alpha-\beta}2}_{B_n}\otimes\ket{\frac{\alpha-\beta}2}_{C_n}\otimes\ket{i}_E d\alpha d\beta\label{multiport output}.}
 From this expression, one can directly see that the reduced density matrix of the output signal-port of Bob is equal to that of Charlie,
 \EQ{\rho_{B_s}=Tr_{C_s,B_n,C_n,E}(\ket{\psi_{out}}\bra{\psi_{out}})=\rho_{C_s}=Tr_{B_s,B_n,C_n,E}(\ket{\psi_{out}}\bra{\psi_{out}}).}

Furthermore, one can see that the resulting state, is symmetric under swap of $B_s$ with $C_s$, which is a property, in general stronger than simply stating that their reduced density matrices are equal. Note, that we have assumed that the initial state of this particular element of the signature we consider here, could have been entangled with any other system, since we gave as initial state a general mixed state (the mixture could have come from tracing out some entangled environment). This is captured by the ``environment'' degree of freedom that appears above as the state $\ket{i}_E$. Moreover, the state of this element of the signature, could have been entangled with other elements of the signature, and the resulting state would still be symmetric under swap of $B_s$ with $C_s$ for \emph{each} element of the signature separately. For example, the two element state \EQ{1/\sqrt{2}(\ket{1}_{B_1}\ket{1}_{B_2}\ket{-1}_{C_1}\ket{-1}_{C_2}+\ket{-1}_{B_1}\ket{-1}_{B_2}\ket{1}_{C_1}\ket{1}_{C_2}),} while the reduce density matrices for $B_1$ and $C_1$ are identical and similarly for $B_2$ and $C_2$, it is not symmetric under swap of subsystems, and thus this state could \emph{not} be the output of the multiport.
Operatively, this entails that, regardless of the prior measurement outcomes Bob and Charlie could have obtained, the unmeasured system is still symmetric under the swaps of Bob and Charlie's signature element pairs.
This stronger property will be important in proving security against coherent repudiation attacks.

The multiport we consider here has two recipients, but it can also be generalized to many recipients~\cite{ErikaOrig}.

\subsection{Security against repudiation}

In a repudiation event, Alice manipulates the quantum signature states \footnote{We assume that all the classical information which is sent from Alice is cross-checked by Charlie and Bob over secure classical channels, so repudiation by tampering with classical messages is not possible.} which are sent in the distribution stage, in such a way that during the messaging stage, the same declaration will be confirmed by one party (checking against the threshold $s_a$) but rejected when forwarded to another party (who checks against the threshold $s_v$).
As we have defined earlier, repudiation is successful, if Bob authenticates, Charlie rejects and there is no abort. To bound on this probability, we can consider the minimum of the following two probabilities, Bob authenticating and Charlie rejecting to verify, i.e.
\EQ{P(rep)\leq P(BA,CR)\leq \min \{P(BA),P(CR)\}\label{repudiation 1}.}
There are two things to note, to understand intuitively why such a bound would be useful. First, we should note that due to the symmetricity of the multiport $P(BA)$ and $P(CR)$ are \emph{not} independent. Second, we should note that the threshold for Bob authenticating is lower than that of Charlie verifying.

We will first consider individual strategies by Alice and then generalize to general separable strategies and to general coherent strategies where we will prove security by reduction to individual strategies. In other words, we will show that the best strategy for Alice is given by individual attacks.

\subsubsection{Security against individual repudiation}

In this type of attacks Alice sends quantum signatures, possibly different, to Bob and Charlie. Each pair of Bob's and Charlie's corresponding quantum signature elements are not correlated with others (classically or through entanglement), but are not necessarily identical, pair to pair. That is, the global state of the quantum signature is in a factorized form, with respect to the partition in different signature element pairs. There may be entanglement within each Bob-Charlie signature element pair.
Because Bob and Charlie pass their quantum signatures through the multiport, the outbound
quantum signature element  at the signal ports of Bob and Charlie is always symmetric under swap of  Bob and Charlie.
That is, the quantum signatures Bob and Charlie collect at their multiport signal ports are such that the reduced density matrix for each element of the signature of Bob is the same as Charlie's.

 Bob's (Charlie's) measured signature agrees/disagrees or neither (ambiguous outcome), with the declared signature for the $i^{th}$ element, with probability $p^i_1,p^i_{-1},p^i_0$. Those probabilities, if the multiport is ideal, are identical for Bob and Charlie. We can define the \emph{average} probabilities $\bar{p}_j=1/L\sum_i p^i_j$, with $j\in\{1,-1,0\}$. Furthermore, we define as $\bar{X}_{-1}$ the empirical (observed) percentage of mismatches over the total number ($L$) of elements of the signature.

It is easy to see that Bob authenticating means that $\bar{X}_{-1}\leq s_a p_{USD}$, which using Eq. (\ref{Hoeffding 2}) leads to
\EQ{P(BA)=P(\bar{p}_{-1}-\bar{X}_{-1}\geq \bar{p}_{-1}-s_a p_{USD})\leq \exp (-2(\bar{p}_{-1}-s_a p_{USD})^2 L)\label{BA},}
provided that $\bar{p}_{-1}\geq s_a p_{USD}+\epsilon'$. Note that $\epsilon'$ is an arbitrarily small positive number. We note that if this is satisfied,
the probability decays exponentially.

Charlie failing to verify means that $\bar{X}_{-1}\geq s_v p_{USD}$ which using Eq. (\ref{Hoeffding 1}) leads to
\EQ{P(CR)=P(\bar{X}_{-1}-\bar{p}_{-1}\geq s_v p_{USD}-\bar{p}_{-1})\leq \exp (-2(s_v p_{USD}-\bar{p}_{-1})^2 L)\label{CR}}
provided that $\bar{p}_{-1}\leq s_v p_{USD}- \epsilon'$. Alice's only freedom, in her attempt to repudiate, is to choose different $p^i_j$'s. In reality, Alice does not have full freedom for these choices (as the POVM elements of Bob and Charlie's measurements do not have orthogonal supports), but since we are looking for a bound, we may assume a worst-case setting and assume that she does have full freedom. Moreover, it is clear from the conditions we have obtained so far that only the averages of $p^i_{-1}$, i.e. $\bar{p}_{-1}$  are relevant for the bound on the repudiation probability we are attempting to establish. Therefore, provided that we choose
\EQ{s_v > s_a\label{Condition 1},}
the probability for repudiation decays exponentially for \emph{all} choices of $\bar{p}_{-1}$. From Eq. (\ref{repudiation 1}), we can see that Alice's best strategy to achieve repudiation is to choose $\bar{p}_{-1}$ in such a way that the minimum of BA and CR is the greatest. From Eq's (\ref{BA}-\ref{CR}) we get that the optimum value is for $\bar{p}_{-1}=\frac{p_{USD}}2 (s_v+s_a)$. This gives the following bound for repudiation under individual attacks,
\EQ{P(rep)\leq \exp (-\frac{p^2_{USD}(s_v -s_a)^2}2 L)\label{Repudiation INID}.}
It is interesting to note that there is an optimal value of
 $\bar{p}_{-1}$ from Alice's point of view, and that her repudiation probability depends on this. Therefore, for the given bound we consider, it is no better for her to choose different values for different $p^i_{-1}$, and thus a strategy where Alice sends identical states for each signature element is at least as effective a full individual strategy for Alice.

\subsubsection{Security against collective and coherent repudiation}

As we have already mentioned, we will prove security against collective and coherent
attacks by showing that those attacks reduce to the individual attacks we already considered.
Note that in USD-based QDS (this may also hold for a broader class of QDS protocols), the success or failure of repudiation depends on the distribution of classical outcomes Bob and Charlie have.
In the proof of security against repudiation against individual attacks, we required the following two assumptions on the distribution of these outcomes:
\begin{enumerate}
\item element independence -- the distribution of outcomes for Bob/Charlie for the $k^{th}$ signature element is independent from other elements
\item Bob/Charlie symmetricity -- the marginal probability distributions of outcomes for the $k^{th}$ element for Charlie and Bob are equal.
\end{enumerate}
The first property is justified by the the type of attack we allow Alice, and the second by the symmetrization property of the multiport. Note that we have not used any assumption on the values of the probabilities for different outcomes.
As mentioned, in practice, if Alice uses individual attacks, not all probability distributions obeying the above two conditions are obtainable (since the measurements are not rank-1 projectors). In particular, she can achieve probabilities distributions that arise from acting at a general single element state ($\rho_{B_i,C_i}$) with the corresponding POVM ($\Pi_j^B\otimes\Pi_k^C$), where $j,k\in\{1,0,-1\}$ and $\{\Pi_j\}_j$ is the three POVM elements corresponding to the USD measurement that Bob and Charlie carry out.
However, the security against individual repudiation was proven without considering this additional restriction on the ability of Alice to manipulate the probabilities. This will mean that we can use the same arguments for security against the most general attacks.

In the following, we will encode the distributions of classical outcomes of Charlie and Bob as quantum states: the outcome $o \in \{-1,0,1 \}$ we denote with the state $\ket{o},$ such that
\EQ{
\bra{o} o' \rangle = \left\lbrace { 0, o \not= o' \atop 1, o=o'} \right.
}
Therefore, any probability distribution of outcomes corresponds to
a mixture of states $\{\ket{o}\bra{o} \}_{o \in \{-1,0,1 \}}$, without loss of generality.

From Alice's point of view (\emph{i.e.} without the knowledge of the obtained outcome), the individual measurements Bob and Charlie perform on their systems are CPTP maps producing mixtures of states $\{ \ket{o}\bra{o} \}_{o \in \{-1,0,1 \}}$, where the probability of the state $\dm{o}$ is given by $Tr(\Pi_o \rho)$, if the input state is $\rho$.
For each signature element, the joint post-measurement state of Bob and Charlie is then of the form
\EQ{
\sigma_{BC} = \sum_{o_B, o_C} p(o_B, o_C) \dm{o_B}_{B} \otimes \dm{o_C}_{C}.
}
In general, the post-measurement state of Bob and Charlie's systems is then a separable state of the form
\EQ{
\rho^{meas} = \sum_{o_B^1, o_C^1, \ldots, o_B^L, o_C^L} p(o_B^1, o_C^1, \ldots, o_B^L, o_C^L) \bigotimes_{k=1}^L \sigma_{B^k,C^k}^{o_B^1, o_C^1, \ldots, o_B^L, o_C^L}.
}
Note that the probability of successful repudiation depends only on the state $\rho^{meas}$.
Moreover, the processes which determines whether repudiation occured or not, can be denoted as a map $\mathcal{R}ep$ which takes $\rho^{meas}$ as input and outputs one bit (encoded in the orthonormal states $\dm{suc}, \dm{fail}$ denoting success or failure of repudiation),
\EQ{
\mathcal{R}ep ( \rho^{meas}) = p^{\rho^{meas}}  \dm{suc} + (1-p^{\rho^{meas}}) \dm{fail}.
}
Since the map $\mathcal{R}ep$ is physical, it is also a linear map. Moreover the success probability of repudiation, given the state $\rho^{meas}$, is given by
\EQ{
Tr \left[ \dm{suc}  \mathcal{R}ep ( \rho^{meas})   \right].
}
But then for the probability of success or failure given a convex combination of input states
\EQ{
\rho^{meas} = \sum_i p(i) \rho^i
}
is given by
\EQ{
Tr \left[ \dm{suc}  \mathcal{R}ep ( \rho^{meas})   \right] &=&  Tr \left[ \dm{suc}  \mathcal{R}ep ( \sum_i p(i) \rho^i )   \right]  \\ &=&\sum_i p(i) Tr \left[ \dm{suc}  \mathcal{R}ep (  \rho^i )   \right] = \sum_i p(i) p^{\rho^i}.
}
Since the last expression above is a convex combination of positive reals, it holds that
\EQ{
 \sum_i p(i) p^{\rho^i} \leq \max_i p^{\rho^i}.
}
That is, given a mixed input state of the form $\rho^{meas} = \sum_i p(i) \rho^i ,$ an upper bound of the repudiation probability is obtained on one of the elements of the mixture $\rho^i$.
Now, applying this convexity argument to the input state of the form
\EQ{
\rho^{meas} = \sum_{o_B^1, o_C^1, \ldots, o_B^L, o_C^L} p(o_B^1, o_C^1, \ldots, o_B^L, o_C^L) \bigotimes_{k=1}^L \sigma_{B^k,C^k}^{o_B^1, o_C^1, \ldots, o_B^L, o_C^L}
}
we see that an upper bound of the repudiation probability can be achieved by considering the input state
\EQ{\bigotimes_{k=1}^L \sigma_{B^k,C^k}^{o_B^1, o_C^1, \ldots, o_B^L, o_C^L},}
for some sequence of outputs $o_B^1, o_C^1, \ldots, o_B^L, o_C^L$.

Assume next that for every input state
\EQ{
\rho^{meas} = \sum_{o_B^1, o_C^1, \ldots, o_B^L, o_C^L} p(o_B^1, o_C^1, \ldots, o_B^L, o_C^L) \bigotimes_{k=1}^L \sigma_{B^k,C^k}^{o_B^1, o_C^1, \ldots, o_B^L, o_C^L}
}
obtained through Charlie's and Bob's measurements as prescribed in our protocol, it holds that for all sequences of outputs $o_B^1, o_C^1, \ldots, o_B^L, o_C^L$ and every $k$, the state
$\sigma_{B^k,C^k}^{o_B^1, o_C^1, \ldots, o_B^L, o_C^L}$ is symmetric under the swap of Bob's and Charlie's systems.
Then, two properties hold:

\begin{enumerate}
\item The probability of repudiation, given the state $\rho^{meas}$, is upper bounded by the repudiation probability given the state $\bigotimes_{k=1}^L \sigma_{B^k,C^k}^{o_B^1, o_C^1, \ldots, o_B^L, o_C^L},
$ for some sequence of outcomes $o_B^1, o_C^1, \ldots, o_B^L, o_C^L$.

\item The probability distributions of outcomes stemming from the state $\bigotimes_{k=1}^L \sigma_{B^k,C^k}^{o_B^1, o_C^1, \ldots, o_B^L, o_C^L}$ satisfy element independence and symmetricity with respect to Bob/Charlie as we discussed previously.
\end{enumerate}
But this means that general repudiation attacks cannot have a higher repudiation probability than individual repudiation attacks. For the latter we have shown that the repudiation probability diminishes exponentially in $L,$ and therefore the same holds for general repudiation attacks.

What remains to be proven is that for all possible states Alice could have prepared, the post-measurement state of Charlie and Bob can be written in the form
\EQ{
\rho^{meas} = \sum_{o_B^1, o_C^1, \ldots, o_B^L, o_C^L} p(o_B^1, o_C^1, \ldots, o_B^L, o_C^L) \bigotimes_{k=1}^L \sigma_{B^k,C^k}^{o_B^1, o_C^1, \ldots, o_B^L, o_C^L},
}
where  for all sequences of outputs $o_B^1, o_C^1, \ldots, o_B^L, o_C^L$ and every $k$, the state
$\sigma_{B^k,C^k}^{o_B^1, o_C^1, \ldots, o_B^L, o_C^L}$ is symmetric under the swap of Bob's and Charlie's systems.
This we now prove.

Let $\rho_{B_1, C_1, \ldots, B_L, C_L}$ be any state of Bob's and Charlie's system, collected at their signal ports, and let $SWAP_k$ be the unitary operator swapping the $k^{th}$ system of Charlie and Bob \footnote{In the real protocol, the systems are measured sequentially, but since the measurements are performed on distinct subsystems they commute, so the order of measurement does not matter.}.
Due to the symmetrization property of the multiport, given by eq. (\ref{multiport output}) it holds that for all $k$,
\EQ{
\rho_{B_1, C_1, \ldots, B_L, C_L}=SWAP_k\left(  \rho_{B_1, C_1, \ldots, B_L, C_L} \right) SWAP_k^\dagger . \label{symm1}
}
After the measurement of the first signature element by both Charlie and Bob the state of the system is
\EQ{
\sum_{o_{B}^1, o^1_{C}} p(o_{B}^1, o^1_{C}) \left( \dm{o_{B}^1} \otimes \dm{o^1_{C}} \right) \otimes
\rho_{B_2, C_2, \ldots, B_L, C_L}^{o_{B}^1, o^1_{C}}.
}
Due to the symmetrization property, we have that
\EQ{
&&\sum_{o_{B}^1, o^1_{C}} p(o_{B}^1, o^1_{C}) \left( \dm{o_{B}^1} \otimes \dm{o^1_{C}} \right) \otimes
\rho_{B_2, C_2, \ldots, B_L, C_L}^{o_{B}^1, o^1_{C}} = \\
&&\sum_{o_{B}^1, o^1_{C}} p(o_{B}^1, o^1_{C}) \left(SWAP_k\left(  \dm{o_{B}^1} \otimes \dm{o^1_{C}} \right) SWAP_k^\dagger \right) \otimes
\rho_{B_2, C_2, \ldots, B_L, C_L}^{o_{B}^1, o^1_{C}}
}
so this state can be written as
\EQ{
\sum_{o_{B}^1, o^1_{C}} p(o_{B}^1, o^1_{C}) \left(1/2 \left(  \dm{o_{B}^1} \otimes \dm{o^1_{C}} + \dm{o_{C}^1} \otimes \dm{o^1_{B}} \right)  \right) \otimes
\rho_{B_2, C_2, \ldots, B_L, C_L}^{o_{B}^1, o^1_{C}}= \\
\sum_{o_{B}^1, o^1_{C}} p(o_{B}^1, o^1_{C}) \sigma_{B_1, C_1}^{o_{B}^1, o^1_{C}} \otimes
\rho_{B_2, C_2, \ldots, B_L, C_L}^{o_{B}^1, o^1_{C}}
}
where the state $\sigma_{B_1, C_1}^{o_{B}^1, o^1_{C}}$ is symmetric under the desired swap.
Crucially, each of the residual, non-measured states $\rho_{B_2, C_2, \ldots, B_L, C_L}^{o_{B}^1, o^1_{C}} $  corresponding to each of the possible measurement outcomes $o_{B}^1, o^1_{C}$ again have the same symmetricity property as given in equation (\ref{symm1}).
To see this, note that the multiport acts independently on each of the $L$ subsystems so the action of the measurement on the first subsystem commutes with the action of the multiport on the second. But then the state of the second subsystem can be imagined to undergo the multiport after the measurement of the first, in which case it is clear the residual states will be symmetric as well, for every outcome. Thus the process can be iterated, obtaining the state of the form
\EQ{
\rho^{meas} = \sum_{o_{B}^1, o^1_{C} , \ldots, o_{B}^L, o^L_{C}} p(o_{B}^1, o^1_{C} , \ldots, o_{B}^L, o^L_{C})  \sigma_{B_1, C_1}^{o_{B}^1, o^1_{C}} \otimes \sigma_{B_2, C_2}^{o_{B}^1, o^1_{C},o_{B}^2, o^2_{C}}  \otimes \cdots \otimes \sigma_{B_L, C_L}^{o_{B}^1, o^1_{C} , \ldots, o_{B}^L, o^L_{C}}.
}
But, by construction, for each $k$ the state
\EQ{
\sigma_{B_k, C_k}^{o_{B}^1, o^1_{C},\ldots o_{B}^k, o^k_{C}}
}
is symmetric under Bob/Charlie subsystem swap.
Thus, the post-measurement state of Charlie and Bob has the desired property, and we have shown that the upper bound on the repudiation probability of Alice for individual attacks, given by Eq. (\ref{Repudiation INID}),  also holds for general attacks.

\subsection{Security against Forging}

As noted in the main text, forging is defined as when a cheating party (Bob) convinces an honest party (Charlie) that Alice had sent a classical message $k$, when Alice has sent no or another meassage.
To do so, the malevolent party needs to guess Alice's declaration. Throughout this section, when we mention declaration, and matching and mismatching with the declaration, this is the ``fake'' declaration of Bob, which is made after Bob uses all his resources to make the best possible
guess. Note also that security against forging as defined also ensures security against message tampering.

There are two natural classifications of forging attack, characterizing the powers available to the forger, and characterizing whether the forger is malevolent from the distribution stage or just during the messaging phase.
For the first classification attack we distinguish individual (which consists of the identical attacks (IID) and the non-identical but uncorrelated (INID)), collective and coherent attacks, as was done for repudiation.
 Concerning the second classification, the forging can be \emph{passive}, where Bob simply attempts to guess the signature in such a way that his declaration gets accepted by Charlie, but is honest during the distribution stage. Otherwise, forging can be \emph{active} where Bob is dishonest also during the distribution stage, where he attempts to modify Charlies quantum signature so that it helps him subsequently during the messaging stage to make Charlie agree with his own declaration.
Here we prove the security against all attacks except the coherent attacks, for both passive and active modes.

From Eq. (\ref{Condition 1}) it is clear that it is easier to attempt to forge a message in the verification stage rather than in the authentication stage. In other words, it is easier to convince Charlie that Bob forwards the message received from Alice, than Charlie convincing Bob that he directly sends the message (pretending to be Alice).

We have defined as successful forging the scenario where Charlie verifies the fake message sent by Bob and there was no abort. To bound the probability for this, we remind the reader that $P(forge)=P(CV,BNA,CNA)\leq P(CV,BNA|CNA)$. By this definition, Bob does not abort no matter what happens in order to succeed in forging.
The probability we use to bound $P(forge)$ is conditional on Charlie not aborting and therefore the number of unambiguous outcomes $K$ is in the interval $[p_{USD}-\delta,p_{USD}+\delta] L$. To put a bound on the forging probability, we assume the worst case scenario, which is that Charlie obtined the lowest possible fraction of  allowed unambiguous outcomes $K=(p_{USD}-\delta)L$,  as in this case it is more likely for Bob to cause a sufficiently small fraction of mismatches, whatever strategy he uses.

\subsubsection{Security against passive collective forging}

In passive attacks, Bob's optimal strategy is defined by a measurement he performs on the quantum signature, which will produce a declaration which has the highest probability of being accepted by Charlie.
By restricting Bob to collective powers, we mean that Bob can perform adaptive measurements, but on individual subsystems only.
In this case, since all the signature elements are independently distributed, it is clear that the optimal measurement for each subsystem cannot depend on prior measurement outcomes, so adaptivity cannot help Bob.

One could wonder whether Bob's best strategy would involve an independent, non-identically distributed (INID) passive attack (\emph{i.e.} differing measurements per elements of the signatures). Bob is successful in forging if he manages to correctly guess a sufficient percentage of signs of the signature. Failure or success can be defined for each element of the signature, \emph{i.e.} if Bob guessed correctly for the $k$'th pulse. It is therefore clear that the best overall strategy would be to repeat, $L$ times (once per element), the best guess for each element, which will be the minimum-error measurement, by definition. Moreover, since all the states he receive are not correlated, it is unlikely any coherent measurement could outperform individual measurements. However, since we for now only consider a forger limited to collective strategies, this is not central to our arguments.

Having established that the best strategy for passive collective attacks is an IID attack with minimum-error measurements, we proceed to bound the probability of success for this type of attacks.
We begin by considering the security of our protocol from passive forging attacks. We define $\bar{Y}_{-1}:= \bar{X}_{-1} (L/K)$ to be the empirically observed percentage of mismatches over \emph{unambiguous} outcomes (note that $\bar{X}_{-1}$ is the percentage of mismatches over the total number of pulses). Bob succeeds in forging, if Charlie, after receiving Bob's (fake) declaration, finds less than $s_v p_{USD}L$ mismatches ($\bar{X}_{-1}\leq s_v p_{USD}$), and thus verifies. We need that
\EQ{\bar{Y}_{-1}\leq s_v \frac{p_{USD}}{p_{USD}-\delta},}
where we have taken the worst case scenario, $K=(p_{USD}-\delta)L$. That is, the smallest $K$ for which Charlie does not abort. Bob can try to guess the correct signature by making a minimum-error measurement on his copy of the quantum signature. The probability of misidentifying is
\EQ{p_{min}=1/2(1-\sqrt{1-e^{-4\alpha^2}}).} Then using Eq. (\ref{Hoeffding 2}) the probability of forging is bounded as
\EQ{P(forge)\leq P\left(p_{min}-\bar{Y}_{-1}\geq p_{min}-s_v \frac{p_{USD}}{p_{USD}-\delta}\right)\leq \exp \left(-2 \left(p_{min}-s_v\frac{p_{USD}}{p_{USD}-\delta}\right)^2(p_{USD}-\delta)L\right),}
provided that $p_{min}\geq s_v \frac{p_{USD}}{p_{USD}-\delta}$. If this holds, we see that the forging probability decays exponentially in the signature length $L$.

\subsubsection{Security against active forging}

For the active forging attacks, we will first consider IID type of attacks where Bob behaves  independently and identically for each quantum signature element.
Then we will show that independent, non-identically distributed strategies (INID) and also collective strategies reduce to, and are not more powerful than, IID strategies.

\subsubsection{IID active forging attacks}

In active strategies, Bob is allowed to alter the quantum signature sent to Charlie during the distribution stage. By altering the quantum signature Charlie receives, Bob tries to increase his probability to later forge a message. However, if Bob sends a different state to the multiport, Charlie's null-port of the multiport will occasionally detect photons. By demanding that the protocol is aborted if many photons are detected at the multiport's null-port, Charlie limits the degree to which Bob can manipulate his quantum signature, reducing Bob's active attacks to (modified) passive.

\begin{figure}
 \includegraphics[width=18cm]{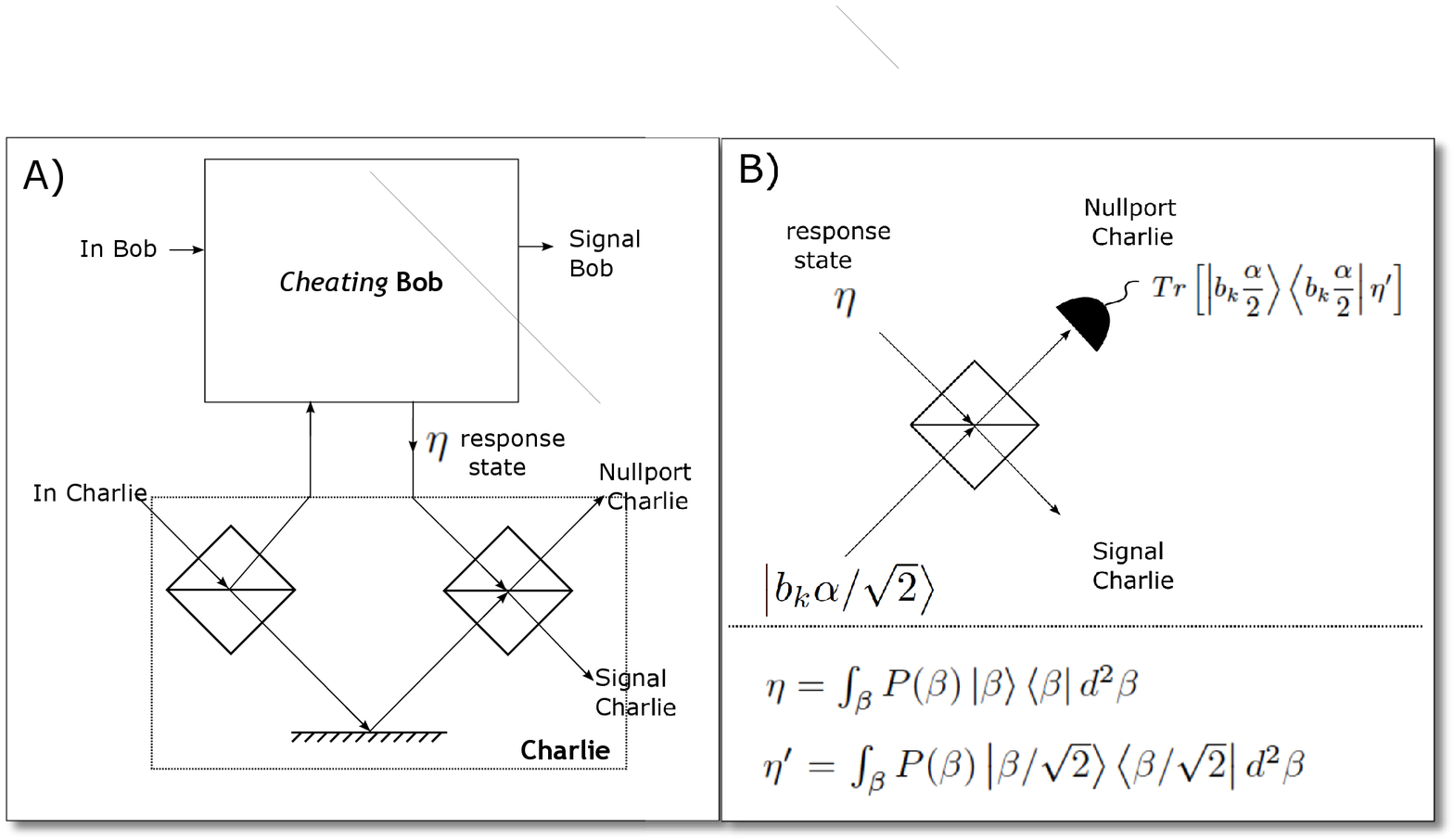}
\caption{\label{Fig:resp} A) In an active cheating strategy, Bob can interfere within the multiport, and alter one of the states Charlie will mix on his final beamsplitter. B) The probability of detecting photons
depends on the response state $\eta$ Charlie received from Bob.  The probability of not seeing a photon at the null-port is equal to the expected fidelity between the honest-setting response state and a 'half' of the response state denoted $\eta'$. }
 \end{figure}

We will call the states that Bob sends to Charlie the \emph{response states}, depicted in Figure \ref{Fig:resp}.
For the $k^{th}$ signature element, let the response state be $\eta = \int_{\beta} P(\beta) \dm{\beta} d^2 \beta$.
Then the probability of obtaining a zero photon count on Charlie's null-port is given by
\EQ{
Tr \left[ \dm{0} \int_{\beta} P(\beta) \dm{\dfrac{\beta}{\sqrt{2}} - b_k \dfrac{\alpha}{2} } d^2\beta \right] = Tr \left[ \dm{ b_k \dfrac{\alpha}{2}} \eta' \right], \label{Dist}
}
where
$ \eta' = \int_\beta P(\beta) \dm{\beta/\sqrt{2}} d^2 \beta$ and $b_k$ is the sign of the $k$'th signature element.
The signal mode is then in the state $ D(b_k \alpha/2)\eta' D^\dagger(b_k \alpha/2)$
where $D(\cdot)$ denotes the displacement operator.

In a passive strategy, the state in the signal mode would be $D(b_k \alpha/2)\eta'' D^\dagger(b_k \alpha/2),$ with $\eta'' = \dm{b_k \alpha/2}$.
Thus, from expression (\ref{Dist}), we can see that the the probability of not detecting a photon at Charlie's multiport null-port is equal to the expected fidelity between the quantum signature elements Charlie will apply USD to in the active and passive attack settings, respectively. This is illustrated in Figure \ref{Fig:resp}. Since we have assumed a separable, identical strategy for Bob for each state, we can use the statistics on this null-port to gauge the distance between the signature states Charlie measures in passive and separable active strategies.
For instance, in the limit of infinite key lengths, if Charlie always observes the vacuum state on his null-port (i.e. detects no photon), Bob must have been behaving honestly within the multiport (barring the possibility of coherent attacks on which we will comment later) so then the expression for the  cheating probability is the same as in the passive setting (and equals zero in the limit of infinite key lengths).

For finite-sized keys, if we can bound the trace distance between the average signal output signature elements in the passive and active attacks, then we can bound the difference between the measurement outcome probabilities of Charlie's USD measurement in the cases of active versus passive attacks.
This is ensured by setting an abort threshold on the multiport null-port photon event count. We will set the threshold to be $r L$, where $r$ is the fraction of the quantum signature states which have caused a photon event at Charlie's null-port during signature distribution, and $L$ is the length of the quantum signature. Recall, we are assuming that Bob is acting independently and identically for each signature element state so this fraction can be used to bound
the value of $Tr(\dm{b_k \alpha/2}\eta')$ for an average signature element state. Let $x:=1-Tr(\dm{b_k\alpha/2}\eta')$. Then, by the Eq. (\ref{Hoeffding 3}) we have that
\EQ{P(|x-r|\geq \epsilon) \leq 2 \exp(-2 \epsilon^2 L).}
Thus, we have that
\EQ{1-Tr(\dm{b_k\alpha}\eta'))\leq r+\epsilon,}
except with probability $2 \exp(-2 \epsilon^2 L)$. In the ideal case, we could set the threshold $r=0$, as we will do at the end when we investigate suitable choices of the parameters. For now, we have $1-Tr(\dm{b_k\alpha/2}\eta'))\leq r+\epsilon$ except
with probability $2 \exp(-2 \epsilon^2 L)$. The variational distance between the probability distribution of a given measurement over an altered state is upper bounded by the trace distance between the original and altered state. To obtain a bound on forging strategies, we need to bound the trace distance. The expected fidelity has a well-known relationship with the trace distance \cite{expfid}, which leads us to a bound on the trace distance,
\EQ{D_{Tr}(\dm{b_k\alpha/2},\eta') \leq \sqrt{1-Tr(\dm{b_k\alpha}\eta')}\leq \sqrt{r+\epsilon}.}
Note that the bound on the trace distance does not hold with probability $2\exp(-2\epsilon^2 L)$. In order to set an upper bound on the forging probability, we will assume the worst case scenario. We will assume that for the cases that the bound does not hold, successful forging had occurred. We will first examine the capacity of Bob to influence states Charlies measures at the end of the distribution stage, assuming the bound on the trace distance holds. This will subsequently affect the matching or mismatching of Charlies quantum signature with Bob's fake declaration, and thus will affect the verification or rejection of Bob's message. To obtain the total forging probability, we will add at the end the probability that the bound on the trace distance does not hold.

Since the number of unambiguous outcomes allowed if abort does not occur
is bounded (more than $(p_{USD}-\delta)L$), and we have already assumed that abort does not occur,
the best strategy for Bob is clearly to try and steer Charlie's outcome towards his declaration, which effectively reduces the percentage of mismatches. The amount he can do this is bounded by $\sqrt{r+\epsilon}$.

Additionally, a cheating Bob has access to the fraction of the state Charlie sends him, giving him a lower minimum error probability should he choose to measure the entirety of the state using a minimum-error measurements. Recall, since all the signs of the quantum signatures are chosen uniformly at random, hence independently from each other, the minimum-error measurement is the optimal strategy for Bob.
The entirety of the system accessible to Bob yields a coherent state of amplitude $\sqrt{3/2} \alpha.$ We will denote the minimum error probability for the minimum-error measurement by $p_{min}'$ and it is given by
\EQ{
p_{min}' = 1 - \dfrac{1}{2}\left( \sqrt{ 1- e^{-6 \alpha^2}} +1\right).
\label{p'min}}
Below, we will give a cheating probability bound which is not tight. We will assume Bob can both measure the entirety of the state, and use it in the preparation of the response state, which is not possible in reality, but upper bounds the cheating probability.

 We repeat the analysis we performed for passive attacks, but with modified values for the constants, (i) instead of $p_{USD}$ we have $p_{USD}-\delta$ which corresponds to the fewer unambiguous outcomes that occur, (ii) instead of $p_{min}$ we have $p'_{min}$ because Bob can use $\sqrt{3/2}\alpha$ for his best guess and (iii) the threshold for verification $s_v$ is effectively increased by the ability of Bob to affect Charlies signature towards his declaration by an amount of $\sqrt{\epsilon+r}$. We then obtain the following expression bounding the cheating probability,
\EQ{P(forge)\leq \exp \left(-2\left(p'_{min}-s_v \frac{p_{USD}}{p_{USD}-\delta}-\sqrt{\epsilon+r}\right)^2(p_{USD}-\delta)L\right).}
As we have already mentioned, the bound on the trace distance we used to derive the expression above holds only with probability $2 \exp(-2 \epsilon^2 L)$. We assume that forging was completely successful for these cases and by the union bound we then have
\EQ{P(forge)\leq \exp \left(-2\left(p'_{min}-s_v \frac{p_{USD}}{p_{USD}-\delta}-\sqrt{\epsilon+r}\right)^2(p_{USD}-\delta)L\right)+2\exp (-2(\epsilon)^2L),\label{active forging}}
provided that
\EQ{p'_{min}-s_v\frac{p_{USD}}{p_{USD}-\delta}-\sqrt{\epsilon+r}>0.}

\subsubsection{INID active forging attacks}

In the section above we considered the setting in which cheating Bob employs an identical strategy for each signature state. Since we could assume that Alice always sends the same signature element state (as this is true up to a local choice of a reference frame of the coherent phase), this implied that the response states Bob sends are all identical (when averaged over the possible outcomes Bob may have obtained internally, which he will later use to make his declaration to Charlie).

Then the null-port photon count directly measured the distance between the honest-case response states, and the one in the IID active attack, which could then be used to bound the overall cheating probability.

In an INID attack, Bob can use a different (but pre-defined) attack per each pulse. This means that the response state depends on the position in the signature, that is, the index of the element.
However, whatever Bob does, his strategy can be fully described with the set of response states he sends,
\EQ{
\eta_{total} = \bigotimes_{k=1}^{L} \eta^k.
}
Multiport counts now gauge the distance between the average response state and the honest response state, where the average response state is given by
\EQ{
\eta_{avg} = 1/L \sum_{k=1}^{L} \eta^k.
}

In the active IID setting, analyzed in the last section, we have used Hoeffding's inequalities to bound the cheating probability. Hoeffding's inequalities give a bound on the probability that the observed average of a sequence of random variables deviates from the expected average by more than some threshold.
However, the only assumption needed for Hoeffding's inequalities to be applicable is that the (binary) variables are independent. This assumption is also satisfied in the case of INID attacks, since the probability distributions in question arise from measurements of a factorized quantum state.
The mean of the realized random variables is directly estimated by estimating the average state Bob sends, \emph{i.e.} the state $\eta_{avg}$.
Thus, the results for the IID attacks directly hold in the non-identical, but independent setting.

\subsubsection{Collective active forging attacks}
In collective forging, Bob's actions on the $k^{th}$ signature element depend on what he did (that is the state of his classical local memory) previously.
An example of such a strategy is one where Bob performs USD measurements up to a certain point.
For instance Bob could do USD continuously (as long as he obtains the unambiguous outcome).
Assume he is lucky, and succeeds in being lucky a sufficient number of times that he is certain he will pass verification. From this point on, it may make sense for Bob to switch to an honest strategy, to reduce the expected average null-port count from this point on.

The key point is, regardless of what happened internally in Bob, what he did could not depend on what happened internally in Charlie, as there is no classical communication between them, due to no-signalling.
Thus, from Charlie's perspective, we can write Bob's cumulative state as
\EQ{
\eta_{tot} = \sum_{i_1, \ldots, i_L} P(i_1, \ldots, i_L) \eta_{i_1} \otimes \cdots \otimes \eta_{i_L}.
}

Next, we can without loss of generality assume that Charlie did not perform any measurements on his system, before he obtained Bob' declaration. The classical system containing the declaration $D$ we will denote $\rho^{D}$.

Now, Charlie's process of verification, which we can denote $\mathcal{V}er$ can be written as a
\EQ{
\mathcal{V}er(\rho^{D} \otimes \eta_{tot}^{D} ) = p^{\eta_{tot}^{D}} \ket{ok}\bra{ok} + \left(1-p^{\eta_{tot}^{D}} \right) \ket{rej} \bra{rej}
}
where the states $\ket{ok}, \ket{rej}$ are orthogonal states of a (qu)bit of Charlie which contains the information whether the message of Bob is verified or not (\textit{rej} denoting rejection).
The probability $p$ depends on  the total input state.
Note that $\mathcal{V}er$ is a deterministic physical process, so a CPTP map, and thus linear.
So we have that
\EQ{
\mathcal{V}er(\rho^{D} \otimes \eta_{tot}^{D} ) = \sum_{i_1, \ldots, i_L} P(i_1, \ldots, i_L) \mathcal{V}er(\rho^{D} \otimes \eta_{i_1}^{D} \otimes \cdots \otimes \eta_{i_L}^{D} ) =\\
\sum_{i_1, \ldots, i_L} P(i_1, \ldots, i_L) \left( p^{\eta_{i_1}^{D} \otimes \cdots \otimes \eta_{i_L}^{D}} \ket{ok}\bra{ok} + \left(1-p^{\eta_{i_1}^{D} \otimes \cdots \otimes \eta_{i_L}^{D}} \right) \ket{rej} \bra{rej}  \right).
}
The probability of cheating is obtained by projecting the state above to the $\ket{ok}$ subspace and  taking the trace, so
\EQ{
P(forge) = Tr \left[\ket{ok}\bra{ok}  \sum_{i_1, \ldots, i_L} P(i_1, \ldots, i_L) \left( p^{\eta_{i_1}^{D} \otimes \cdots \otimes \eta_{i_L}^{D}} \ket{ok}\bra{ok} + \left(1-p^{\eta_{i_1}^{D} \otimes \cdots \otimes \eta_{i_L}^{D}} \right) \ket{rej} \bra{rej}  \right)
\right].}
Since everything above is linear, we have
\EQ{
P(forge) = \sum_{i_1, \ldots, i_L} P(i_1, \ldots, i_L)  p^{\eta_{i_1}^{D} \otimes \cdots \otimes \eta_{i_L}^{D}} .}
Thus the cheating probability is just a convex combination of the cheating probabilities of the individual states, and is maximized for one of the factorized states.

\subsection{Correctness of the protocol}

To confirm the correctness of the protocol, we must check that if all parties behave honestly, then Bob will authenticate and Charlie will verify the forwarded message and neither will abort. Here we are considering the ideal case, where the multiport is perfect and that there are no losses or noise. The non-ideal, experimentally implementable case, is work in progress.

In the honest, ideal case, neither Bob nor Charlie ever rejects a message. However, they may abort. In particular, as mentioned in Section I, there are two types of abort. In the honest, ideal case, Charlie and Bob will never detect any photons at their null-port of the multiport, and thus the only reason to abort that can occur is if they receive fewer or more than expected unambiguous outcomes. This happens if Bob (and similarly Charlie) find that they have unambiguous outcomes outside the interval $[(p_{USD}-\delta)L,(p_{USD}+\delta)L]$ which according to Eq. (\ref{Hoeffding 3}) leads to the probability of honest abort (due to, say, Bob aborting)
\EQ{P(\textrm{honest abort Bob})=P(|1-\bar{X}_0|\geq \delta) \leq 2\exp (-2\delta^2 L).}
The probability of Bob not aborting in the honest case is $[1-P(\textrm{honest abort Bob})]$ and similarly for Charlie. Therefore the probability for neither of them aborting is bounded from below by
\EQ{P(\textrm{honest not abort})=P(\textrm{correct})\geq \left(1-2\exp(-2\delta^2 L)\right)^2.}
We can see that this approaches unit probability exponentially quickly in the signature length $L$.

\subsection{Constraints and Choices of Parameters}

In this section we review the main expressions and constraints on the parameters that we have obtained from correctness, security against repudiation and forging. We will also suggest a suitable choice for the parameters. The probabilities of correct execution, repudiation and forging are bounded by the expressions
\EQ{P(\textrm{correct})&\geq& \left(1-2\exp(-2\delta^2 L)\right)^2\label{eq1}\\
P(rep)&\leq& \exp (-\frac{p^2_{USD}(s_v -s_a)^2}2 L)\label{eq2}\\
P(forge)&\leq& \exp \left(-2\left(p'_{min}-s_v \frac{p_{USD}}{p_{USD}-\delta}-\sqrt{\epsilon+r}\right)^2(p_{USD}-\delta)L\right)+2\exp (-2(\epsilon)^2L)\label{eq3}.}
Moreover we have the  two constraints
\EQ{s_v&>&s_a\label{eq4}\\
p'_{min}-s_v\frac{p_{USD}}{p_{USD}-\delta}-\sqrt{\epsilon+r}&>&0\label{eq5}.}
The parameters that appear are $\delta, s_v,s_a,r,\epsilon$, and $p'_{min},p_{USD}$, the latter being functions of $\alpha$ which is the amplitude of the coherent states. Since we are dealing with the ideal case, we can make the following two simplifications. We can set $r=0$, i.e. demand that there are no photons detected at the null-port of the multiport, and furthermore, we can set $s_a=0$, in other words authenticate \emph{only} if there is not a single mismatch. Note however, that these simplifications are not desirable
in the non-ideal case, because it will greatly increase the probability of abort even when all participants are honest,  and lead to low robustness of the protocol.

From its definition, we can see that $\delta$, the unambiguous count tolerance should be function of $p_{USD}$, while from Eq. (\ref{eq3}) we see that $s_v$ and $\epsilon$ should be functions of $p'_{min}$. One possible choice which guarantees that the protocol is both correct and secure is $\delta=0.1 p_{USD}$, and $s_v=\sqrt{\epsilon}=p'_{min}/4$. We then obtain the simplified expressions
\EQ{P(cor)&\geq& \left(1-2\exp(-0.02p^2_{USD}L)\right)^2\\
P(rep)&\leq& \exp(-0.03 (p_{USD}p'_{min})^2L)\\
P(forge)&\leq& \exp(-0.4 p'^2_{min}p_{USD}L)+2\exp(-0.008 p'^4_{min}L).}
We can easily see that the probability of correctness if all parties are honest approaches unity exponentially quickly, and the probabilities of repudiation and forging tend to zero exponentially quickly.

Finally, one may wonder what would be a suitable choice of amplitude $\alpha$. When $\alpha$ tends to zero then $p_{USD}$ will tend zero, whereas when $\alpha$ tends to infinity, then $p'_{min}$ tends to zero, and neither is satisfactory. The first corresponds to very low intensity, where there are almost no unambiguous outcomes. The second corresponds to very high intensity, and in this case it becomes easy to forge, because we are now in the classical regime where a forger can guess the state correctly with high probability, \emph{i.e.} $p'_{min}\rightarrow 0$. A suitable choice that simnultaneously optimizes  all expressions considered here is difficult to find analytically. However, with the choices made above, taking $\alpha\sim 0.2$ is a reasonable choice. With this indicative choice of $\alpha$ we obtain $p'_{min}=0.27$ and $p_{USD}=0.077$. We can see that if for example $L=10^6$ (which can be practical if high repetition rate pulse lasers ($>$100MHz) are used to generate the coherent states), the probability for correct protocol is very close to unity $P(\textrm{correct})\geq 1-10^{-52}$, while the repudiation probability is very close to zero $P(rep)\leq 2\times 10^{-6}$ and similarly the forging probability $P(forge)\leq 0.7\times 10^{-18}$ (using the bounds we derived in this paper which are not tight).

\end{document}